\documentclass[12pt]{article}

\begin{document}

\begin{flushright}
hep-ph/0609095
\end{flushright}
\vskip .4cm
\begin{center}
{\Large \bf Lepton flavour violation in  Little Higgs model with T-Parity }
\vskip 1cm
Ashok Goyal \footnote{E--mail :agoyal@iucaa.ernet.in}\\ 
{\em Department of Physics and Astrophysics,University of
Delhi, Delhi-110 007, India}\\
{\em Inter-University Centre for Astronomy and Astrophysics,
Pune-410 007, India}
\end{center}

\begin{abstract}
If neutrino mass and mixing consistent with the neutrino oscillation
data are the only source of lepton flavor violation (LFV)in nature, the
other LFV decays like the radiative and semileptonic decays would be
too small to be observed experimentally in the foreseeable
future. These decays have been the objects of recent Belle
measurments. We analyze LFV in Little Higgs Model with T-parity and
find that with reasonable values of the model parameters, these decays
can very well be experimentally observable. 
\end{abstract}

\vskip 1.5cm
\begin{section}{Introduction}
 The neutrino oscillation data  from the Solar,
Atmospheric and Accelerator experiments presents a compelling evidence for the 
existence of small neutrino mass and large neutrino flavour mixing. The SK
 atmospheric neutrino and K2K data \cite{one} are best described by dominant
$\nu_{\mu}\rightarrow\nu_{\tau}$ vacuum oscillations with best fit
values $|{\Delta M_A}|^2 = 2.1\times10^{-3}eV^2$ and
$sin^2{2\theta_A}=1.0$ at $99.73\%$ CL. The Solar neutrino data
is described by $\nu_e\rightarrow\nu_{\mu}$ oscillations with best fit
value ${\vert{\Delta M_0}}^2\vert = 7.9^{+0.6}_{-0.5}\times10^{-5}eV^2$ and
$tan^2{\theta_0}=0.40^{+.09}_{-.07}$. The Troitzk and Mainz tritium
$\beta$-decay experiments \cite{two} provide information on the absolute
$\bar{\nu_e}$ mass measurment $m_{\bar{\nu_e}}<2.2$ eV at
$95\%$ CL. From the study of anisotropy in the CMBR and large
 scale structure, the WMAP data \cite{three} has severe constraints on the masses
of all active neutrino species $\sum m_{\nu_j}<(0.7-1.8)eV (95\%$
CL).This shows that neutrino flavour is not conserved in nature. In the
 minimal Standard Model(SM)which has been remarkably successfull in explaining
all electro-weak symmetric (EWS)interactions probed so far, the neutrinos are
 massless because of the restricted particle spectrum and requirement of
 gauge invariance and renormalizability. One could easily accomodate neutrino
 mass in the SM by introducing a right handed state resulting in a Dirac mass
term through Yukawa coupling just as for charged fermions. This Yukawa
 coupling of course , has to be roughly six orders of magnitude smaller
 than for charged fermions in order to obtain small neutrino masses consistent
 with the above data. This feature is generally considered unnatural. If small
 neutrino masses in SM are the only source of lepton flavour violation
 (LFV), other LFV processes like radiative lepton decays
 ($\mu \rightarrow e \gamma$ etc.), semileptonic decays 
($\tau \rightarrow \mu M$ ) and trileptonic decays
 ($\tau \rightarrow e^-(\mu^-)\mu^+ \mu^- $) which are objects of recent
 Belle measurments \cite{four}, would be so supressed by the small neutrino mass and
 leptonic GIM mechanism that these 
observatiions in the foreseeable future are well-nigh impossible.
If we simply introduce a neutrino mass in the SM consistent with experiments
and a GIM type lepton flavour mixing matrix, the branching ratio of radiative
lepton flavour violating decay $\mu \rightarrow e \gamma$ has a value less 
than ${10}^{-40}$. Thus there is a need to explore other sources of LFV in
theories beyond the SM.
 
 Neutrino mass generation is not the only problem
afflecting SM. The other problem is the so called Hierarchy problem,
that is enormous difference between the electro-weak and GUT/Planck
scale. The precision electro-weak data prefers the existence of light
Higgs and thus SM with light Higgs can be considered as an effective
theory valid to a high scale perhaps all the way to GUT/Planck scale
whereas the Higgs mass is not protected and gets quadratically
divergent contribution to its mass and requires fine tuning. Supersymmetry
is one of the most attractive framework  wherein the  quadratic
divergences contribution to  Higgs mass is cancelled between particles of 
different statistics at the 1 TeV scale. However with the prospects of LHC
 drawing near which will test supersymmetry, there have been recently 
alternative approaches to address this problem. One of the approaches
 popularly known as Little Higgs Model\cite{five} treats  Higgs
fields as Nambu-Goldstone bosons of a Global symmetry which is
spontaneously broken at some high scale f by acquiring vacuum
expectation value (vev). The Higgs field gets a mass through
electro-weak symmetry breaking triggered by radiative corrections
leading to Coleman-Weinberg type of potential. Since the Higgs is
protected by approximate Global symmetry, it remains light and the
quadratic divergent contributions to its mass are cancelled between
particles of the same statistics. The Littlest-Higgs (LH) model is a
minimal model of this genre which accomplishes this task of cancelling
quadratic divergence to one loop order with a minimal matter
content. The LH model consists of an SU(5) non-linear sigma model
which is broken down to SO(5) by a vacuum expectation value f. The
gauged subgroup $[SU(2)\times U(1)]^2$ is broken at the same time to
diagonal electr-weak SM subgroup $SU(2)\times U(1)$. The new heavy states in
this model consist of vector 'top quark' which cancels the quadratic
divergence coming from the SM top quark along with the new heavy gauge
bosons $(W_H,Z_H,A_H)$ and a triplet Higgs $\Phi$, all of masses of
order f and in the TeV range. The effect of these new states on
electro-weak precision parameters has been studied to put constraints
on the parameters of the model \cite{six}. However, the precision electro-weak
 observables due to the exchange of heavy degrees of freedom in the model
get contributions at the tree level. This requires the cutoff scale of 
new physics to be $\sim 5-10$ TeV and reintroduces the hierarchy
problem \cite{six}.

Motivated by these constraints, a new implementation\cite{seven} of the LH model has
 been proposed. This is done by invoking a discrete symmetry called the
T-parity in the model. T-parity explicitly forbids any tree level contribution
 from heavy gauge bosons in the e.w. precision observables. It also forbids the
interaction that induces the triplet vev. As a result the corrections to
e.w. precision observables are generated at the one loop level only. It makes 
the constraints much weaker than in the tree level case and fine tuning is 
avoided. In this model SM particles are even under T-parity and most 
of the new particles at the TeV scale including the Higgs triplet $\Phi$
 are odd. Another attractive feature of this model is that the lightest 
T-odd particle ($A_H$ is neutral and if T-parity is conserved, can be a 
 candidate for the dark matter (WIMP) much like the neutralino in MSSM 
with R-parity.

The particle content of LH model with T-parity consists of \cite{eight}

\begin{itemize}
 
\item[1] T-odd partners of SM gauge bosons $W_H, Z_H$ and $A_H$ with masses 
$M_{W_H} = M_{Z_H} = gf , M_{A_H} = g\prime f/ \sqrt 5$.

\item[2] T-odd partners of SM fermions (quarks and leptons) with masses 
typically $\sim \sqrt 2 \kappa f $ of TeV order.

\item[3] A triplet $\Phi$ of T-odd Higgs with masses $M_{\Phi}^2 = \frac
{2M_H^2f^2}{v^2}$

\item[4] A vector T-odd and a singlet T-even top quark with T-odd partner
 being lighter than the T-even partner both masses being in the TeV range.

\item[5] In addition there is a T-odd doublet $\tilde \Psi_R$ and a 
singlet $\xi_R$ which are needed to cancel two loop quadratic divergence
 to the Higgs mass but otherwise are assumed to decouple from the spectrum.
 Their masses are much larger than the symmetry breaking scale 
 ($\sim  5f$ ) and they have negligible effect on the low energy
phenomenology but at the same time their masses are low enough to
 keep the Higgs mass small.   
\end{itemize}

In this model the T-odd heavy gauge bosons have gauge interactions with
 the T-odd heavy fermions and T-even SM fermions.The interaction that generates
masses of T-odd fermions also couples the T-odd scalar triplet $\Phi$
to SM fermions through $\tilde \Psi_R$ through the Yukawa coupling. These 
interactions can be extended to include generation mixing through CKM type
matrices just as in SM .The generations would now mix and unless we have a 
universally degenerate mass spectrum for the T-odd fermions, the interaction
 would result in FCNC and LFV \cite{nine}.

In the notation of Hubisz et.al\cite{eight}, the interactions are given by
\begin{equation}
  {\cal L}_G=g \bar \Psi_{Hi}V^{\dag i}_{Hj} G_H V_{SMk}^j \Psi_{SM}^k  +hc
\end{equation}
\begin{equation}
{\cal L}_Y = \kappa^i_j f (\bar \Psi_{2i} \xi \tilde \Psi^j + \bar \Psi_{1i}
\Sigma_0\Omega\xi^{\dag}\Omega \tilde \Psi^j) + hc
\end{equation}

where the rotation matrices are related to the appropriate CKM matrix
as in SM and $\Psi_{SM}$ and $\Psi_H$ are T even and odd fermion doublets.
\end{section}
\begin{section}{ Lepton Flavour Violation}

Radiative leptonic decays $\mu\rightarrow e\gamma,\tau\rightarrow e \gamma$
 along with semileptonic decays $\tau\rightarrow\mu\pi(\eta,\eta\prime, K)$
 which are objects of recent Belle measurments \cite{four} are sensitive
 probes of LFV. In order to obtain information on LFV couplings we have to
 confront anomalous magnetic moment of the muon which is by far the 
most precisely measured quantity in nature.

{\bf a) Anamolous magnetic moment of  muons}

  Theoretical predictions of SM 
for $a_{\mu}= \frac{g-2}{2}$ with experimental results \cite{ten} give
\begin{equation}
a_{\mu}(E821) - a_{\mu}(SM) = (25.2-26.0\pm9.4)\times 10^{-10}
\end{equation}

In LH model with T-parity the contributions to $a_{\mu}$ comes from the 
exchange of heavy vector bosons and Higgs triplet $\Phi$. We have calculated
 the contribution from the exchange of these particles in the Unitary 
gauge and obtain 
\begin{equation}
a_{\mu}(W_H)= -\frac{g^2}{32\pi^2}\frac{m_{\mu}^2}{M_{W_H}^2}\Sigma V_
{H2i}^*V_{Hi2}F_{W_H}\left[Z_i=\left(\frac{M_{\nu_{Hi}}}{M_{W_H}}\right)^2\right]
\end{equation}
\begin{equation}
a_{\mu}(Z_H)= \frac{g^2}{32\pi^2}\frac{m_{\mu}^2}{M_{Z_H}^2}\Sigma V_
{H2i}^*V_{Hi2}F_{Z_H}\left[Z_i=\left(\frac{M_{l_{Hi}}}{M_{W_H}}\right)^2\right]
\end{equation}
\begin{equation}
a_{\mu}(A_H)=a_{\mu}(Z_H)[g\rightarrow g\prime /5 , M_{Z_H}\rightarrow M_{A_H}]
\end{equation}

where
\begin{equation}
F_{W_H}(Z_i) = \frac{1}{6(Z_i-1)^4}[-10+37Z_i-48Z_i^2+7Z_i^3+14Z_i^4+
(-12Z_i^4-6Z_i^3+24Z_i^2)lnZ_i]
\end{equation}
\begin{equation}
F_{Z_H}(Z_i) = \frac{1}{12(Z_i-1)^4}[-8+38Z_i-39Z_i^2+14Z_i^3-5Z_i^4+
18Z_i^2)lnZ_i]
\end{equation}

Contribution of $\Phi$ involves exchange of mirror leptons ($\tilde L$) which 
are supposed to decouple

\begin{equation}
a_{\mu}(\Phi^-)= \frac{\kappa^2}{32\pi^2}\frac{m_{\mu}^2}{M_{\Phi}^2}
\Sigma V_{H2i}^*V_{Hi2}F_{\Phi}\left[Z_i=\left(\frac{M_{\tilde \nu_{Hi}}}{M_{\Phi}}\right)^2\right]
\end{equation}
\begin{equation}
F_{\Phi}(Z_i) = \frac{1}{6(Z_i-1)^4}[-1-3Z_i^2-2Z_i^3+6Z_i^4+
6Z_i^2)lnZ_i]
\end{equation}

From eqns. (4) and (5) in the limit of small neutrino mass , we get the
SM contribution to $a_{\mu}$ given by
\begin{equation}
a_{\mu}(SM)=\frac{g^2}{48\pi^2}\frac{m_{\mu}^2}{M_{W_L}^2}
\lbrace \frac{5}{2}-(1+2S_W^2-4S_W^4)\rbrace
\end{equation}

For representative values of masses of T-odd particles discussed
above, we find that the contribution of T-odd vector bosons, fermions
and scalars is not more than a few percent of the SM contribution and
therefore well within control.

{\bf b) Radiative decays}

As discussed above if small neutrino mass and large mixing as required by 
neutrino oscillation data are accomodated in SM by simply introducing small
neutrino mass, other LFV process $\mu \rightarrow e \gamma$ will have a 
branching ratio $<10^{-40}$ so small that there would be no hope of detecting
this in the foreseeble future. In LH model with T-parity there is a possibility 
of branching ratio being enhanced even with a TeV level GIM mechanism. Present
experimental limits on radiative decays of leptons at 90 $\%$ CL \cite{four}
are 

\begin{equation}
 Br(\mu \rightarrow e\gamma)<1.2\times 10^{-11}, 
 Br(\tau \rightarrow \mu \gamma)<6.8\times 10^{-8}, 
 Br(\tau \rightarrow e\gamma)<3.92\times 10^{-7}
\end{equation}

 Branching ratios can be easily calculated and we get 
\begin{equation}
BR(W_H) = \frac{3}{2}\frac{\alpha}{\pi}\left(\frac{M_{W_L}}{M_{W_H}}\right)^4
\delta_{W_H}^2 = 8.15\times 10^{-7}\delta_{W_H}^2
\end{equation}

\begin{equation}
BR(Z_H) = \frac{3}{8}\frac{\alpha}{\pi}\left(\frac{M_{W_L}}{M_{Z_H}}\right)^4
\delta_{Z_H}^2 = 2.04\times 10^{-7}\delta_{Z_H}^2
\end{equation}
\begin{equation}
BR(A_H) = \frac{3}{200}(\frac{g`}{g})^4\frac{\alpha}{\pi}\left(\frac{M_{W_L}}
{M_{A_H}}\right)^4\delta_{A_H}^2 = 2.57\times 10^{-7}\delta_{A_H}^2
\end{equation}
\begin{equation}
BR(\Phi) = \frac{3 \alpha}{\pi}(\frac{\kappa}{g})^4\frac{M_{W_L}^4}{\tilde
 M_{\nu_H}^2 m_{\mu}^2}
\delta_{\Phi}^2 = 8.15\times 10^{-7}\delta_{\Phi}^2
\end{equation}
In SM we have
\begin{equation}
BR(SM) = \frac{3 \alpha}{32 \pi}\delta_{\nu}^2 <10^{-40}
\end{equation}
where

\begin{equation}
\delta_{SM}= \Sigma V_{\mu i}^* V_{ie}(\frac{m_{\nu i}}{M_{W_L}})^2
\end{equation}
\begin{equation}
\delta_{V}= \Sigma V_{H\mu i}^* V_{Hie} F_V(Z_V)
\end{equation}
\begin{equation}
F_{W_H}(Z_i) = \frac{1}{12(Z_i-1)^4}[10-43Z_i+78Z_i^2-49Z_i^3+4Z_i^4+
18Z_i^3lnZ_i]
\end{equation}
\begin{equation}
F_{Z_H}(Z_i) = \frac{1}{12(Z_i-1)^4}[-8+38Z_i-39Z_i^2+14Z_i^3-5Z_i^4+
18Z_i^2)lnZ_i]
\end{equation}
\begin{equation}
F_{\Phi}(Z_i) = \frac{1}{6(Z_i-1)^3}[-1+Z_i^2-2Z_i lnZ_i]
\end{equation}

Substituting these in the branching ratio  expressions, we find that
if order one mixing angles are allowed in the heavy fermion sector , a
Tev scale GIM suppression is necessary and even a mass spectrum which
is degenerate upto few percent is enough to make the branching ratios
accessible to the present experimental limits \cite{four}

{\bf c) Semi-leptonic decays}

Recent experimental searches by Belle \cite{four} give
\begin{equation}
BR(\tau \rightarrow\mu\pi)<4.1\times10^{-7},BR(\tau \rightarrow\mu\eta)<1.5\times10^{-7},
BR(\tau \rightarrow\mu\eta\prime)<4.7\times10^{-7}
\end{equation}

 In the LH model with T-parity, we get new contributions to the
semileptonic LFV decays. These contributions come from Box diagrams
that contain T-odd heavy gauge bosons, heavy quarks and leptons. The
LFV arises mainly because of generation mixing in the heavy leptonic
sector. The dominant contribution to the LFV effective Hamiltonian can
be estimated following $\Delta S=2$ effective Hamiltonian calculations
for $K^0-\bar K^0$ mixing \cite{eleven}. The Hamiltonian has V-A structure
and the leading term is proportional to $\frac{v^2}{f^2}$ and can be
written as 
\begin{equation}
 {\cal H}_{eff} = \frac{G_F^2}{64\pi^2}M_{W_L}^2\frac{v^2}{f^2}\sum
\lambda_i\lambda_j A_{ij}(Z_i,Z_j)(\bar d d)_{V-A}(\bar\mu\tau)_{V-A}
\end{equation}
which can be written as 
\begin{equation}
{\cal H}_{eff} = \frac{1}{\lambda^2}(\bar d d)_{V-A}(\bar\mu\tau)_{V-A}
\end{equation}
where
\begin{equation}
\frac{1}{\Lambda^2}\simeq 0.8\times 10^{-10}\sum
\lambda_i\lambda_j A_{ij}(Z_i,Z_j) GeV^{-2}
\end{equation}

The decay widths can now be easily calculated and we get 
\begin{equation}
\Gamma(\tau\rightarrow\mu\pi) \simeq 0.9\times 10^{-18}\vert \sum
\lambda_i\lambda_j A_{ij}\vert ^2 GeV
\end{equation}
and 
\begin{equation}
BR(\tau\rightarrow\mu\pi) \simeq 0.25\times 10^{-7}\vert \sum
\lambda_i\lambda_j A_{ij}\vert ^2 
\end{equation}

Comparing with the experimental data we see that in the LH model with
T-parity there is a real possibility of observing these LFV
semi-leptonic decays at the present experimental sensitivity for
realistic values of generation mixing and even for small departure
from the mass degeneracy in the heavy fermion sector.

\end{section}
\begin{section}{Conclusions}

Little Higgs Model with T-parity is an attractive framework which
successfully adresses the hierarchy problem and makes the electro-weak
precision constraints much weaker. This happens essentially because in
this model e.w. precision observables are generated only at the one
loop level. The model has the added attraction of providing a possible
candidate for dark matter (WIMP)in the form of lightest T-odd neutral
gauge boson $A_H$ much like neutralino in MSSM with R-parity. The
model can be easily extended to include generation mixing in the heavy
fermion sector and can give FCNC and LFV interactions at the
observable level. If order one mixing angles are allowed in the heavy
fermion sector, a TeV scale GIM supression is necessary and a few
percent of departure from mass degeneracy in the heavy lepton mass
spectrum is enough to make the lepton flavor violating radiative and
semileptonic decays experimentally accessible in the near future.
\end{section}

\section*{Acknowledgments}
I thank the organisers of VI Rencontres du Vietnam where this work was
presented. I thank S.R.Choudhury
and Naveen Gaur for helpful discussions.

\end{document}